\documentstyle[12pt]{article}
\input amssym.def
\input amssym.tex
\begin{document}
\begin{center}
{\bf TOPOLOGY AND TIME REVERSAL}\\
\vspace*{1cm}
{A. CHAMBLIN \& G.W. GIBBONS}\\
{D.A.M.T.P.}\\
{University of Cambridge}\\
{Silver Street}\\
{Cambridge CB3 9EW}\\
{U.K.}\\
\vspace*{1.5cm}

{\bf ABSTRACT}\\
\end{center}

{\small
In this lecture we address some topological
questions connected with the existence on a
general spacetime manifold of diffeomorphisms
connected to the identity which reverse the time-orientation.}

\section {Introduction}

If one regards Quantum Gravity as an attempt to unify two
distinct but equally fundamental physical theories;
quantum mechanics on the one hand and general relativity
on the other, one can ask what elements of either theory
is it most likely that one will have to sacrifice in the eventual
unification. Perhaps the most fundamental innovations
of general relativity relate to its treatment of the notion
of time. One of most striking features of quantum mechanics is
its use of  complex amplitudes. One may argue that the introduction
of complex numbers into the basic structure of quantum mechanics
is closely connected to the treatment in that theory of the notion
of change and of time evolution. It therefore seems reasonable to
regard the use of complex numbers in conventional
quantum mechanics as a potential casuality. More precisely,
one may argue that if, as is commonly supposed in quantum cosmology,
 the classical idea of time is an emergent concept,
valid only at late times, low energies and large distances, then so too
is our usual idea of a quantum mechanical  Hilbert space
with its attendant  complex structure. {\it In other words, the complex numbers
in quantum mechanics should be thought of as having
an essentially historical origin}. Some ideas along these lines
were discussed within the context of the semi-classical approach
to quantum cosmology in [16--17].

A related question is to ask:
how in a theory in which one assumes that spacetime has an
everywhere  well-defined  Lorentzian metric are the properties
of quantum fields in those spacetimes affected by such global properties
of the spacetime as the existence of closed timelike curves (\lq CTCs \rq),
a lack of time-orientability or some other pathology
which would normally be excluded in a globally hyperbolic spacetime?
Are there restrictions on the possible spacetimes for example?
One possible restriction comes about by demanding that spacetime admit a
spin or pin structure [6].
Another possible restriction arises by demanding that
 the spacetime has a
time-orientation. If it does not, one may argue that one
may not be able to construct a quantum mechanical Hilbert space endowed with a
complex structure. This suggestion was made some time ago [25]
and it has received further
support from the work of Bernard Kay [7].

One motivation for asking this question is to try to extend the
range of applicability of quantum field theory in a fixed background.
Another motivation might be to answer questions about what possibilities
the laws of physics in principle allow. This has provided much of the
impetus behind recent work on CTCs . Another, and possibly more cogent,
 reason for considering
non-globally-hyperbolic spacetimes is that in the path integral approach
to quantum gravity in which one sums over all possible
Lorentzian metrics
there is {\it a priori} no good reason for excluding them.
One might attempt to perform the functional integral by first freezing the
metric and integrating over all matter fields on that spacetime, and then
summing over all spacetimes. The first part of the integral is then
tantamount to quantizing matter fields on a fixed background.
It is customary in the Euclidean formulation
 to replace the sum over Lorentzian metrics by a sum
over Riemannian metrics but one may ask what happens
if one tries to avoid this step.

In the Euclidean version one is often concerned with
anomalies that may  arise when functional determinants fail to be well-defined,
for example they may not be invariant under spacetime
diffeomorphisms. The diffeomorphisms in question may either be
continuously connected to the identity or not. The latter type of
global anomalies are closely related to discrete
symmetries, or lack of them, such as parity or orientation.
They may also be investigated from an Hamiltonian point of view .
However, this does not address the possibility of anomalies
of a purely Lorentzian kind which manifest themselves
only in non-globally-hyperbolic spacetimes.
An example is a breakdown of spin structure.
If one assumes that spacetime is both time and space-orientable, this
can {\it only} occur in a spacetime which is not globally hyperbolic .
If one drops the requirement of space-orientability, however,
 there may exist  pin structure even though the spacetime {\it is}
globally hyperbolic. An example is provided
by ${\Bbb R}{\Bbb P}^2 \times {\Bbb R}^2$, endowed with the product metric
formed from the standard `round' metric on ${\Bbb R}{\Bbb P}^2$
and the Minkowski metric on ${\Bbb R}^2$ (with either signature) [6].

One possible viewpoint on the difficulties experienced with
non-time-orientable spacetimes is precisely that there is some
sort of anomaly. Roughly speaking, for each  complex amplitude
in the functional sum one must, if there is no global time-orientation,
include its complex conjugate  which is associated to the time-reversed
amplitude. The result must then necessarily be real and so no true
quantum interference is possible. It is interesting to note
that this sort of problem would also arise in some
attempts to generalize
the usual quantum formalism being made by Gell-Mann and Hartle [22] since they
also
make use of
{\it complex} amplitudes and they incorporate a rule relating
complex conjugation to time reversal of a sequence of observables.

The purpose of this lecture is to explore
some of these issues in more depth.
In particular, we will discuss the relation between the
topology of a time-orientable  spacetime $\{{\cal M}, g\}$
and the existence and properties of various kinds
of time-reversing diffeomorphisms.
We shall, for the sake of mathematical
precision,
mainly concentrate on spacetime manifolds ${\cal M}$ which are compact and
without boundary,
but we will comment on the case of non-compact spacetimes and spacetimes
with boundaries.

As well as the motivations given above, our results are
also relevant to  suggestions like that of Sakharov [18] that
the early universe may simply be a time reflection
of the late universe.  Such a viewpoint is essentially a Lorentzian version
of the (historically later) no-boundary proposal
or the idea of a universe born from nothing [23].\\

\section {Compact Spacetimes}

An assumption of compactness in spatial directions is
quite natural when discussing topological questions because one has in mind
a situation where the non-trivial topology can be localized, at least to the
extent that it is not allowed to escape from the spacetime altogether.
Compactness in the time direction is less easy to justify
(unless there are spacelike boundaries) because it necessarily implies the
existence of closed timelike curves . Formerly this was thought
to rule out consideration of such spacetimes but more recently,
with the advent of studies of the properties of time machines,
this view has been abandoned and so we shall not be put off by this feature.

In fact the Euler number $\chi({\cal M})$ of a compact spacetime of
arbitrary dimension
must vanish:
\begin{equation}
\chi({\cal M}) = 0,
\end{equation}
and in four dimensions:
\begin{equation}
\chi = 2 -2 b_1 +  b_2,
\end{equation}
where $b_i$ are the Betti numbers. Thus a compact spacetime must have an
even second Betti number and infinite fundamental group, and so its universal
covering space is
non-compact. In this sense it may be thought of as a non-compact spacetime
which has been periodically identified and this is indeed typically
how
examples of time machines are constructed in the literature.
However, the reader is cautioned that
there is, as we shall see later, no
logical connection between whether or not a curve
is closed and timelike and whether or not it is homotopically trivial.
In general, one expects the fundamantal group $\pi_1({\cal M})$
to be non-Abelian. This is what one expects in the case
of two or more  time machines, for example if the spacetime
has in a connected sum decomposition two summands of the form
$S^1 \times S^3$ with time running around the $S^1$ factors.

In the exceptional case that the fundamental group {\it is} Abelian,
it may be shown [8--11] that  the possible Betti numbers $(b_1,b_2)$
must belong to the set: $\{$(1,0), (2,2),
(3,4), (4,6)$\}$ .
This is because for any closed orientable manifold of any dimension
which has an Abelian fundamental group one has the inequality:
\begin{equation}
{ 1 \over 2} ~b_1 ( b_1-1) \le b_2
\end{equation}

The result follows from (1) which holds for any spacetime dimension
and (2) which holds in four dimensions.

The  significance of this non-Abelian-ness in the case that
homotopically non-trivial time machines
are present is presumably that
some physical effects may depend upon the order in which one
enters the time machines. It would be interesting to
explore this point further. In that connection, it is perhaps
worth recalling why it is that
non-simply-connected four-manifolds are not classifiable [26].
The point is that by taking the connected sum $\# _k S^1 \times S^3$
 of  $k$ copies of
$S^1 \times S^3$ one obtains a four-manifold whose fundamental group
is the free group on $k$ generators (which of course is
maximally non-Abelian). One may now perform surgery on this manifold
to obtain a new manifold whose fundamemtal group
has $k$  generators and $r$ arbitrarily chosen relations.
Since there is no algorithm for deciding whether two
different presentations give
an isomorphic group there can be no algorithm for deciding whether two
four-manifolds are homeomorphic .

The process of surgery can be described as follows.
Given an element $g \in \pi_1({\cal M}^\prime)$
of a four-manifold $\cal M^\prime$ one can represent it by a closed
curve $\gamma \in {\cal M}^\prime$. Now surround this closed curve
$\gamma$ by a tube or collared neighbourhood ${\cal N}$ of the form
${\cal N} =D^3 \times \gamma \equiv D^3 \times S^1$ where $D^3$ is a closed
3-dimensional disc. The boundary $\partial {\cal N}$
 of this tube has topology $\partial {\cal N} \equiv S^1 \times S^2$.
One now removes the tube ${\cal N}$ from ${\cal M}^\prime$
and replaces it with the
simply connected  manifold $D^2 \times S^2$ which has the same boundary.
The result is a new manifold ${{\cal M} ^ \prime} ^ \prime$ whose
fundamental group differs from that of ${\cal M}^\prime$ only by the
imposition of the relation $g= 0$. This process is called
\lq killing an element of the fundamental group\rq.
It may be shown that by a succesion of such killings one may obtain from
$ \#_k S^1 \times S^3$ a manifold with any desired
finitely generated fundamental group.

{}From a physical point of view it is interesting to note
two things. Firstly that the undecidability problem
reviewed above may give
rise to limitations on what is \lq in principle\rq  ~allowed by the
laws of physics when it comes to
the sort of wormhole and time machine engineering
envisaged by Thorne and others. The possibility arises of
having two sets of instructions for building a multiple
time machine
but having no algorithm for deciding whether the two spacetimes have the same
topology. Whether or not this is true is not obvious from
the general result quoted above because  a compact spacetime
must have vanishing Euler number. We do not know whether such
manifolds are classifiable or not.

The second point is that the process of surgery gives
rise to a manifold which  physically looks rather like
one containing the creation and annihilation of an extra
Einstein-Rosen throat. If the 2-disc $D^2$ has coordinates
$X+iT= r \exp \left(it- {{i\pi} \over 2}\right)$ where the cyclic \lq time\rq
{}~coordinate
$t$ which parameterizes the original curve $\gamma$
runs between $0$ and $2\pi$ then \lq half-way round\rq,  i.e.
on the real axis $T=0$,  the interior of the tube ${\cal N}$
has been replaced by a manifold which has the same topology
as the Kruskal manifold of a black hole and therefore it
has embedded in it a three-manifold which has the topology
of a bridge, i.e. of ${\Bbb R}\times S^2$. If
these sorts of manifolds do arise in a Lorentzian
form of quantum gravity it seems reasonable to think
of them as containing \lq virtual black holes\rq.

This interpretation receives some support from the  observation that the
Riemannian manifolds used
as instantons or real tunnelling geometries in the
Euclidean approach to vacuum instability and black hole pair
creation may be  obtained by surgery on a circle
, which we would like to associate with the world line of a virtual black hole,
from the corresponding false vacuum spacetime.  Thus the Euclidean
Schwarzschild manifold ($ {\Bbb R}^2 \times S^2$)
may be obtained from the hot flat space manifold $S^1 \times {\Bbb R}^3$,
the Ernst instanton manifold
($S^2 \times S^2 - \{pt\}$)   for the creation of
pairs of oppositely charged non-extreme
black holes from a constant electromagnetic field (topology ${\Bbb R}^4$), and
the Nariai and Mellor-Moss Instantons (both with  topology
$S^2 \times S^2$) are obtained  from the De Sitter manifold
 ($S^4$). In Kaluza-Klein theory, Witten [27] has argued that the
five-dimensional Schwarzschild solution (topology ${\Bbb R}^2 \times S^3$)
is the bounce solution which mediates the decay of the Kaluza-Klein
vacuum (topology $S^1 \times {\Bbb R}^4$ ).
The five-dimensional manifold corresponding to a
magnetic field also has topology $S^1 \times {\Bbb R}^4$.
This may decay via Witten's instability but it may also decay into
a monopole-anti-monopole pair. The instanton for this process has topology
$S^5-S^1 \equiv {\Bbb R}^2 \times S^3$ and so may also be obtained by surgery
on a circle from the false vacuum spacetime manifold.\\

\section {Time Reversal in a General Spacetime}

Let $\{ {\cal M},g^L \}$ be a time-orientable spacetime.
Thus the bundle of time-oriented frames $SO_{\uparrow}(n-1,1) ({\cal M}, g^L)
$ falls into two connected components.
One typically thinks of time reversal $\Theta$ as
a diffeomorphism:
\[
\Theta: {\cal M} \rightarrow {\cal M}
\]
which reverses time-orientation, whose lift to
$SO_{\uparrow}(n-1,1) ({\cal M}, g^L)
$  exchanges the two connected components
 and is an involution of order two:
\[
\Theta ^2 ={\rm id}.
\]
It need not necessarily be an isometry (in general the spacetime
will not admit any isometries). One could imagine considering
a more general finite group action but presumably one could
always find a ${\Bbb Z}_2$ subgroup and we shall assume that
this can be done.

In a general non-globally-hyperbolic spacetime
it is not obvious whether
$\Theta$ should reverse space-orientation,
or total orientation (assuming $\{{\cal M}, g^L \} $ to
be space or time-orientable respectively) , whether it should act freely
on ${\cal M}$ or fix a three-surface for example, or whether
it should belong to the identity component ${\rm Diff} _0 ({\cal M})$.
The existence and uniqueness and other properties
of $\Theta$   depends
both on the  topology of the manifold $\cal M$  and on the Lorentz
metric $g^L$.

To illustrate these subtleties,
consider even the simplest  globally hyperbolic spacetime
${\cal M} \equiv {\Bbb R}
\times \Sigma$ with coordinates $t,{\bf x}$, $t$ being timelike
and $\Sigma$ being an orientable $(n - 1)$-manifold. Naively we might take
\[
\Theta ^T: (t, {\bf x}) \rightarrow (-t, {\bf x})
\]
but nothing prevents us from considering
\[
\Theta ^J : (t, {\bf x}) \rightarrow (-t, {\bf x}^*)
\]
where
\[
J : {\bf x} \rightarrow  {\bf x}^*
\]
is an involution on the $(n-1)$-manifold $\Sigma$. Clearly $\Theta ^T$
fixes the three-manifold $\Sigma$ and reverses total orientation.
It therefore lies outside the identity component ${\rm Diff}_0({\cal M})$.
On the other hand, we might arrange for $J$ to act freely
on $\Sigma$, possibly reversing or not reversing space-orientation.

These seemingly rather artificial examples actually arise
in some applications. In quantum field theory in De Sitter spacetime,
${\rm dS}_n$,
$\Sigma$ is the $(n - 1)$-sphere and $J$ its antipodal map.
This preserves space-orientation if the spacetime dimension $n$ is
even. The map $\Theta ^J$ is an isometry and is the
centre of the isometry group $O(n,1)$. One
may identify points under the action of $\Theta ^J$
to obtain the \lq elliptic interpretation\rq. This
then provides a possible non-singular realisation
of Sakharov's ideas of a Lorentzian model
of a universe born from nothing. The idea immediately generalizes to
a Friedman model whose scale factor is an even function of time.
Sakharov's idea was in fact to impose some sort of time-reflection
symmetry about a singular big bang at which the scale factor vanishes.
He did not use the involution $J$.
In spatially closed
 models
the scale factor often starts from a zero value at the big bang, $t=0$,
rises to a maximum at $t=t_{\rm max}$ say, and then
symmetrically decreases to a vanishing value at the  big crunch at $t=2t_{\rm
max}$.
This has led Gold [2] to  conjecture that the \lq arrow of time reverses\rq
{}~in the contracting phase. In effect  he proposed
that the entire quantum state is invariant under a time-reversing
involution
whose action on spacetime is given by:
\[
\Theta ^G: ~t \rightarrow t_{\rm max} -t.
\]
By contrast Davies [1] (see also Albrow [4])
prefers to  continue  through the Big Bang and Big Crunch to get a
model in which the arrow of time reverses in sucessive cycles.
In other words, one imposes invariance under the action of
semi-direct product ${\Bbb Z} \odot {\Bbb Z}_2$ given by
\[
t \rightarrow t+2 t_{\rm max}
\]
and
\[
t \rightarrow  -t.
\]
It is clear that similar options are available for non-singular
periodic models in which there is neither a Big Bang nor a Big Crunch.
Thus for example, in the case of
Anti-De Sitter spacetime ${\rm AdeS}_n$,
the scale factor is a sinusoidal function of
cosmic time but the vanishing of the scale factor is an
artefact of a poor choice of coordinates. In fact
${\cal M} \equiv S^1 \times H^{n-1}$ where $H^{n-1} \equiv {\Bbb R} ^{n-1}$
is hyperbolic space and time $t$ runs around the circle, $0\le t <2 \pi$.
The center of the isometry group $O(n-1,2)$
does not reverse time (it sends $(t,{\bf x})$ to $(t+ \pi ,-{\bf x})$).
Intuitively, it seems clear that time reversal must have fixed points since
we must reverse $t$ and compose with an involution $J$
which may be thought to act on Euclidean space.

In the examples so far (at least if  we wish to maintain
the boundary conditions) there was  no  natural choice of $\Theta$
in the identity component ${\rm Diff}_0( {\cal M})$.
However in more exotic situations,
as we shall see in detail shortly, this seemingly
paradoxical situation can occur.
Now if no  possible $\Theta$ lies in the
identity component ${\rm Diff}_0( {\cal M})$
it is reasonable to say that the spacetime $\{{\cal M}, g^L\}$
has an intrinsic sense of the passage of time
(even though time itself may not be defined!).
If however there exists a $\Theta$ which does lie in the identity component
this is not reasonable.
The general situation with respect to ${\rm Diff}({\cal M})$
appears to be quite difficult
to analyse and so we shall restrict attention here
to a simpler question.
Is there a homotopy rather than a diffeomorphism carrying the metric
$g^L$ with one time-orientation to the same metric with the opposite
time-orientation? If there does exist a suitable $\Theta$ in the identity
component ${\rm Diff} _0({\cal M})$ then a homotopy will certainly
exist
(simply pull back $g^L$  by a curve $f_s$, $0\le s \le 1$
 of diffeomorphisms joining $f_1= \Theta$ to the identity $f_0= {\rm id}$).
However the converse is not necessarily true. Given a homotopy $g_t^L$
of Lorentz metrics
there may exist no diffeomorphism producing it.
Now from the point of view of homotopy theory,
a closed
time-oriented Lorentzian spacetime $\{ {\cal M}, g^L\} $
contains  no more information than a Riemannian manfold ${\cal M}$
equipped with a unit vector field ${\bf V}$. The spacetime with the
opposite time-orientation corresponds homotopically
to the same manifold equipped with the negative unit vector field $-{\bf V}$.\\

\section {Mathematical Interlude}

This following mathematical interlude  follows  some conversations with Graeme
Segal.\\

\subsection {Linear and General Homotopies}

We suppose that $M$ is a closed, n-dimensional
time-orientable Lorentzian manifold.
We may, in the standard way, endow $M$ with a Riemannian
metric and hence deduce that $M$ admits a global section ${\bf V}$ of
the bundle $S(M)$ of unit vectors over $M$. At each point
$x$ in $M$ the fibre $S_x$ of $S(M)$ is an $n-1$ sphere.

Pulling the Lorenzian metric back under the action of diffeomorphisms
induces an action on $\bf V$ and we would like to know whether
there exists a diffeomorphism $f: M \rightarrow M$ which takes ${\bf V}$
to its negative, i.e. which reverses the direction of time.
In particular we would like to know whether there exits such a
diffeomorphism $f$ contained in the identity component
${\rm Diff}_0(M)$ of the diffeomorphism group
${\rm Diff}(M)$. An easier question to ask is
whether there exists a homotopy
taking ${\bf V}$ to $-{\bf V}$ since if there exists a
diffomorphism in the identity component a homotopy
is given by a curve $f_t$ in ${\rm Diff}_0(M)$
joining $f$ to the identity.
The converse is however
 not sufficient because, as we shall see, if one considers
$M=S^1 \times S^{2n-1}$ with the vector field running around the
 $S^1$ factor one finds that
this cannot be reversed by a diffeomorphism but it may be
reversed by a homotopy

A homotopy ${\bf V}_t$
between ${\bf V}$ and $-{\bf V}$ thus gives at each point $x$ in $M$  a
continuous path $\gamma_x(t)$ from the north pole to the south pole of
$S^{n-1}$.
In other words a {\sl general homotopy} ${\bf V}_t$
provides a global section $s_Z$ of a bundle $Z(M)$ whose
fibres
$Z_x$ are the space of paths from the north to the south pole of $S^{n-1}$.
Since any path from the north pole to the south pole of $S^{n-1}$ is
homotopic to a closed  path on $S^{n-1}$ one sees that
 from the point of view of  homotopy the fibre $Z_x$ is equivalent to
the loop space $\Omega (S^{n-1})$ of based loops on $ S^{n-1}$ .

Consider now a {\sl special} or {\sl linear} homotopy from
$\bf V$ to $-{\bf V}$.
By definition this is one for which, at each point $x$ in
$M$, ${\bf V(x) }_t$ lies in a
an oriented two plane $\pi(x)$ spanned say by the vectors ${\bf V}_0$ and
${\bf V}_{t_1}$ where $0 < t_1 < 1$.
A linear homotopy gives a
particular kind of path $\gamma_x(t)$ from the north to the
south pole of $S^{n-1}$,
one which is along a great circle in the 2-plane  defined by by the vectors
${\bf V}_0$ and ${\bf V}_{t_1}$. The set of
such great circles is parameterized by where the great circle
intersects the equatorial $S^{n-1}$.

The existence of
a linear homotopy is thus equivalent to the existence of a global
section $s_Y$ of the $S^{n-2}$ bundle  $Y(M)$ of
unit vectors orthogonal to the vector field $V(x)$.
One may think of this $S^{n-2}$ fibre $Y_x$ as the equatorial $S^{n-2}$
in the $S^{n-1}$ fibre $S_x$  of the bundle $S(M)$.
 It follows that
the bundle $Y(M)$ is a sub-bundle of the bundle $Z(M)$.
The question of whether every homotopy can be deformed into a
linear homotopy then
reduces to the question whether every section $s_Z$ may be deformed
to a section $s_Y$.

It should also be clear that the existence of the vector field
and a linear homotopy is
equivalent to a non-vanishing section of the bundle $ V_{n,2}(M)$
of dyads, i.e of ordered pairs of linearly independent vectors
${\bf e}_1$ and ${\bf e}_2 $ say. The fibre of the dyad bundle
$V_{n.2}(M)$ is
the Stiefel manifold $V_{n,2}$  of dyads. In addition a linear homotopy
provides a global section $s_G$ of the bundle $G_{n,2}(M)$
of oriented 2-planes whose fibre is
the Grassman manifold $G_{n,2}$. The existence of a section $s_G$
is, in fact,  the necessary and suffient condition  that a manifold admit a
metric of signature $(n-2,2)$.

We note {\it en passant} the following

\noindent {\bf Lemma} { \it  If $M$ is even dimensional
a sufficient condition for $M$ to admit a linear homotopy is that it admit
 an almost complex structure $J$. In four dimensions this condition is also
neccessary.}

The point is that  one may then  take
$$
{\bf V}_t= e^{t\pi J} {\bf V_0}
$$
In four dimensions the existence of an almost complex structure
is also necessary since given the
dyad field
one obtains an almost  complex structure by extending the rotation
through $\pi \over 2$  in the  two-plane spanned by the two vectors
 to the unique orthogonal two-plane. The sign ambiguity may be fixed by
the convention that the associated two-form is anti-self-dual.

A simple example is provided by the manifold mentioned earlier:
$S^1 \times S^{2n-1} $.
As is well known this is a complex manifold and hence it certainly admits a
complex structure.
Thus the vector field which just winds around the $S^1$ factor can certainly
be reversed by a homotopy but it is clear,
by using a metric to convert the vector to a one-form and considering
the line integral of the one-from around the circle,   that it cannnot  be
reversed by
 a diffeomorphism. To see that $S^1 \times S^{2n-1} $ is a complex manifold
one notes that $S^1 \times S^{2n-1} \equiv {\Bbb C}^{2n} /{\Bbb Z}$ where the
integers
${\Bbb Z}$ act on ${\Bbb C}^{2n} \equiv {\Bbb R}^{4n}$ by
$(z^1, z^2, \dots, z^n) \rightarrow
(\lambda ^m z^1, \lambda ^m z^2, \dots \lambda ^m z^n,)$
where $m \in {\Bbb Z}$ and $\lambda $
is a real number not equal to zero or unity.

Now it is known, eg. from Morse theory,  that the homotopy
type of the fibre $Y_x$ of loops on $S^{ n-1}$ is that of a cell-complex
corresponding to the geodesic paths. Thus there is a cell corresponding
to going once around the sphere, the descending directions parmetrizerized
by the equatorial
$S^{n-1}$, and next comes a cell $S^{2(n-2)}$ and so-forth.
If $n>4$ this second cell is higher in dimension than the dimension of the base
$M$
of the bundle $Y(M)$. It follows from obstruction theory  that there
is no obstruction to pushing points of any  section $s_Y$ in the fibre $Y_x$
down onto the $S^{n-2}$ of the fibre $Y_x$ of the dyad bundle $Y(M)$.
In other words we have the following

\noindent {\bf Proposition}:  {\it In dimensions greater than $4$ a general
homotopy is
deformable to a linear homotopy and thus the necessary and sufficient
condition for a general homotopy is the existence
a global section of the $S^{n-2}$  bundle $Y(M)$, or equivalently a global
section of the dyad bundle $V_{n,2}(M)$.}

Atiyah [15] (see also Thomas [14])  has  obtained some necessary conditions for
the existence of
a non-singular dyad field. From their work one has one has the following

\noindent {\bf Proposition}  {\it
A necessary condition that a $4 k$ dimensional manifold, $k>1$
admit a time revesing homotopy is that
the signature $\tau(M)$ be divisible by $4$. A necessary condition that a
$4k+1$
dimensional manifold admit a time-reversing homotopy is that the real Kervaire
semi-characteristic $k(M)$ vanish. }

The real Kervaire semi-characteristic is defined by
$$
k(M) = \sum  b_{2p} \>{\rm mod}\>  2
$$
where $b_{2p}$ are the Betti numbers, $b_{2p} = {\rm dim} \> H^{2p} (M; {\Bbb
R})$.  \\

\subsection {Four-dimensional case}

In four dimensions the situation is  more delicate because the
cell-decomposition of the fibre $Y_x$ contains  a  4-sphere which has
the same dimension as the base.
We now turn to a more detailed discussion
of the four dimensional case.
We begin with some general facts about $S^2$ and $S^3$ bundles
over oriented four manifolds.

Firstly note that oriented 3-plane, or equivalently
$S^2$ bundles $Q \rightarrow M$
have characteristic  classes $ w_2 \in H^2(M;{\Bbb Z}/2)$
 and $p_1 \in H^4( M; {\Bbb Z})$ which satisfy:
\[
p_1 = w_2^2 \> {\rm mod } \> 4
\]
The characteristic classes $w_2$ and $p_1$ subject to this condition
determine and are
determined by the bundle $Q$. Moreover, the bundle $Q$ admits a cross section
if
and only if
there exists an element $\xi \in H^2(M; {\Bbb Z})$ such that
\[
\xi = w_2  \> {\rm mod} \> 2
\]
and
\[
\xi^2 = p_1.
\]

The class $\xi$ may be thought of as follows. A non-zero section $s$ of an
oriented three-plane  bundle gives rise  to an oriented
2-plane bundle whose fibres consist of vectors orthogonal
to the section $s$. This oriented two-plane bundle may be thought of
as a complex line bundle and $\xi$ is its first
Chern class $c_1$.

Similarly, in four dimensions, real four-dimensional oriented vector
bundles $E \rightarrow M$
determine and are determined by classes $w_2 \in H^2 (M; {\Bbb Z}/2)$
and $p_1, e \in H^4(M;{\Bbb Z}) $ such that
\[
w_2^2 = p_1 + 2 e  \> {\rm mod} \> 4.
\]
Given $E$ one may pass to the bundle of two forms
$\Lambda (E)$.  Giving the fibres a positive definite metric
we  obtain two 3-plane bundles, $\Lambda ^{\pm}$
of self-dual or anti-self-dual two forms. We have
\[
w_2 (E) = w_2(\Lambda ^+) = w_2 ( \Lambda ^-)
\]
and
\[
p_1(\Lambda ^\pm) = p_1(E) \pm 2 e(E)
\]
We are of course interested in the case when $E$ is the tangent bundle
of the manifold $M$. Then
$w_2$ is the second Steifel-Whitney class $w_2(M)$,
$e$ its Euler class $e(M)$, and $p_1$ its Pontryagin class $p_1(M)$.
The Pontryagin class is related to the signature $\tau$ by
\[
p_1(M) = 3 \tau (M)  ,
\]
moreover
\[
\tau = e \> {\rm mod} \>2
\]

Now if $E$ admits a global section $s_E$
(which can happen if and only if the
euler class $e$ vanishes) then the bundles
$\Lambda^+(E)$ and $\Lambda^-(E)$ are isomorphic. This is because
given any vector $u$ orthogonal to the section $s_E$ we get a
self or anti-self  dual two form, i.e.
\[
u \wedge s_E \pm \star u \wedge  s_E.
\]
Thus both $\Lambda ^+$ and $\Lambda ^-$ are isomorphic to the
bundle of vectors orthogonal to $s_E$, $E^\perp$. The set of such
unit vectors corresponds to the equatorial two-sphere in the three-sphere
in our general discussion above.

  An almost complex
structure  or equivalently a linear homotopy therefore exists if and only if
there exists a section $u_{E^\perp} $. Such a section exits if and only if
there exists an an element $\xi \in H^2(M; {\Bbb Z})$ such that
\[
\xi = w_2 \>{\rm mod } \>2
\]
and
\[
\xi ^2 = p_1=3 \tau.
\]
Now in four dimensions $F^2 (M) =H^2(M;{\Bbb Z})/ {\rm Tor} (M) $
is an integral lattice since it is equipped, via the cup product, with an
integral valued bilinear product: the intersection form $Q(\>,\>)$. By Wu's
formula
the second Stiefel Whitney class satisfies
\[
Q(w_2,x) =Q(x,x)
\]
for all $x\in F^2 (M)$ and thus
\[
Q(\xi, x)= Q(x,x) \>{\rm mod} \>2
\]
for all $x \in F^2(M)$, in other words the element $\xi$ is
a so-called {\sl characteristic} element of the integral lattice
$F^2(M)$. It follows on purely arithmetic grounds, by  a lemma of Van der Blij
[24], that for such an element
\[
Q(\xi,\xi) = \tau \>{\rm mod}\> 8.
\]
The other condition on $\xi$ becomes
\[
\xi^2 = p_1= 3 \tau
\]
that is, eliminating $\xi^2$
\[
2 \tau = 0 \>{\rm mod }\>8
\]
or
\[
\tau = 0 \>{\rm mod }\>4.
\]

\noindent{\bf Lemma} {\it A neccessary condition that a closed Lorentz
4-manifold admit a linear homotopy is that the signature is divisible by 4}

Moreover we have also shown that

\noindent {\bf Proposition } { \it A Lorentz 4-manifold admits a linear
homotopy if
and only if it admits an almost complex structure }

These conditions are non-trivial because, while for a spin manifold $\tau = 0
\,{\rm
mod}\,16$ [14], in general one only knows that if the Euler characteristic
nanishes then  $\tau = 0 \>{\rm mod}\>2$. In fact
to obtain an example of a Lorentz 4-manifold which does not admit a linear
homotopy
consider  the connected sum of $2n$ copies of ${\Bbb C} {\Bbb P}^2$
with $n+1$ copies of  $S^1 \times S^3 $. This has $\tau =2n$, and does not
admit a spin structure, even if $n$ is a multiple of $8$.
Unless $n$ is divisible by $4$ it cannot admit a linear homotopy.

We may relate this discussion to the
question of the existence of global sections of the bundle of dyads
$V_{4,2}(M)$, a subject studied by Hirzebruch and Hopf [28].
Generically a section will have singularities
isolated at points in the manifold $M$. Surounding each
point by small 3-sphere we get a map from $S^3 \rightarrow V_{4,2}$.
Since $\pi_3(V_4,2) \equiv {\Bbb Z} \oplus {\Bbb Z}$ one has
an index consisting of two  integers $(a,b)$ associated
with each singularity. If $\alpha = Q (\xi,\xi)$ where $\xi$ is a
characteristic element of
$H^2(M; {\Bbb Z})$ then the allowed values  are given by
\[
(a,b) = { 1 \over 4} ( \alpha - 3 \tau -2 e, \alpha - 3 \tau + 2 e).
\]

In terms of Betti numbers one has
\[
{ 1\over 4} (3 \tau +2 e) = { 1 \over 4} (b^+-b^-) +  1 -b_1
\]
and
\[
{ 1 \over 4} ( 3 \tau -2 e) = { 5 \over 4} (b^+ -b^-) -1+b_1
\]
So the integrality of $(a,b)$ is automatic. In the present case $e=0$ and if we
have a global section then there exists
a characteristic element $\xi \in H^2(M; {\Bbb Z})$ such that

\[
 \xi ^2 = 3 \tau
\]
 This is of course the same  condition that we used above.

Our necessary  condition for the existence of an almost complex structure may
also be
obtained by considering the index of the associated Dolbeault complex. This is
called the arithmetic genus, $ag(M)$. In dimension 4
\[
ag(M) = {1 \over 4} \,(\chi + \tau)= { 1 \over 2} (b^+ + 1 - b_1)
\]
and thus under our assumption that $\chi = 0$ this again leads to
the necessary condition for the existence of
an almost complex structure is that the signature $\tau$ be divisible by 4.

Consider the more general problem of whether a general homotopy exists.
This requires the existence of a section of the bundle $Z(M)$. As always,
the potential obstructions lie in $H^i(M; \pi_{i-1}( Z_x))$.
Since $M$ is four dimensional $H^i(M; \pi_{i-1}( Z_x))$ for $i>4$,
so we need only consider $\pi_i(Z_x)$ for $i\le 3$. Now $Z_x$ is the
space of based loops on $S^3$ and so
\[
\pi_i(Z_x)=\pi _{i+1} (S^3).
\]
The possible obstructions are therefore in $H^i(M; \pi_i(S^3))$.
Thus there are two potential obstructions: the primary one,
which is an element of  $H^3(M; {\Bbb Z})$
and a secondary one which is an element of $H^4(M;{\Bbb Z}/2)$.

The primary
obstruction coincides with the obstruction for the bundle
$Y(M)$ and is the third integral Stiefel Whitney class $W_3 (M)$
which is the obstruction to lifting the second
${\Bbb Z}/2$-Stiefel Whitney class $w_2 \in H^2(M;{\Bbb Z}/2)$
to an integral class $\xi$. It is the obstruction to the introduction of a
${\rm Spin}_c$ structure,
$\xi$ being the Chern class of the circle bundle..
This is well known to vanish for an orientable four-manifold.
There remains  the secondary obstruction. In the case of the bundle
$Y(M)$ this vanishes if $\xi ^2= p_1$. In the case of the bundle $Z(M)$
it   vanishes under the weaker
condition that
\[
\xi ^2 = p_1 = 3 \tau \> {\rm mod}\> 8
\]

But as before $\xi^2$ is congruent to $\tau$ mod $8$ and thus we have the
following

\noindent {\bf Proposition} {\it
The
necessary and sufficient condition for general homotopy is
\[
 \tau = 0 \, {\rm mod } \>4.
\]}

The necessary and sufficent condition for a general homotopy is the same as the
necessary condition for a linear homotopy obtained above. It is
a non-trivial requirement as the examples constructed above illustrate.
The remaining
question, whose answer is not known at present,  is whether the
necessary condition is sufficient. This boils down
to a purely arithmetic question about the possible intersection forms $Q$.  \\

\section {Some Examples}

Every odd dimensional
sphere $S^{2r+1}$ admits a time-orientable
Lorentz metric $g^L$. One takes:
\[
g^L= g^R - 2 {\bf V}^\flat \otimes {\bf V}^\flat,
\]
where $g^R$ is the standard round metric, ${\bf V}^\flat$ is the one-form dual
to
the vector field ${\bf V}$ obtained by using the musical isomorphism (i.e.
lowering the index with the metric $g^R$) and the unit vector field ${\bf V}$
is
tangent to the Hopf fibration. If $Z^a$, $a=1,\dots 2r+2$
 are complex coordinates
for $ {\Bbb R}^{2r+2} \equiv {\Bbb C}^{r+1}$ then the Hopf fibration
corresponds to the
$SO(2) \subset SO(2r+2)$ action :
\[
Z^a \rightarrow \exp (it) Z^a
\]
and
\[
{\bf V}= {\partial \over {\partial t}}.
\]
The case $r=1$ should be familiar because it is
encountered in the Taub-NUT solutions of
Einstein's equations. The general case also arises in higher dimensions
as we shall describe later.

 Atiyah's result tells us that if $r$ is even, $r=2k$, then
the Lorentz structure described above cannot be obtained by
a diffeomorphism which is connected to the identity
 to the
Lorentz structure
whose light cones differ merely by being upside down.
On the other hand, we may  trivially reverse
the light cones by using the diffeomorphism $\Gamma$ consisting
of $r$ reflections $\in O(2r+2)$, i.e. by complex conjugation:
\[
\Gamma : \hskip 0.5cm Z^a \rightarrow {\bar Z^{\dot a}}.
\]

Now if $r$ is odd then $\Gamma$ lies in the identity component
$SO(2r+2)$ of $O(2r+2)$ and hence in the identity
component  ${\rm Diff}_0(S^{2r+1})$ of ${\rm Diff}(S^{2r+1})$.
If however $r$ is even, $r = 2k$, then $\Gamma$ is not in the identity
component
of $O(4k+2)$ and, by Atiyah's result, not in ${\rm Diff}_0(S^{4k+2})$ either.

Thus Lorentz metrics on $S^{4k+1}$ of the type we have been considering fall
into
two classes with opposite time-orientation. This
is similar to the situation with respect to orientation
(i.e. combined space and time-orientation).  An oriented manifold
may or may not be diffeomorphic to the same manifold with the opposite
orientation. Manifolds which are, are called {\it reversible}.
Manifolds which are not, are called {\it irreversible} or
sometimes {\it chiral}. Of course in this latter case
the diffeomorphism must
lie outside the identity component ${\rm Diff}_0({\cal M})$.

Chiral manifolds are analogous to enantiomorphic crystal forms,
such as seen in quartz for example. In that case, they arise
because the point group
of the crystal is contained in  $SO(3)$ and thus
includes no orientation reversing isometries of Euclidean space.
In our case,
however, we are {\it not} requiring our diffeomorphism
 to be an isometry of any metric.

The spheres $S^n$ are obviously reversible
because they admit reflections.
By contrast, some of the three-dimensional lens spaces, $L_{p,q}$ (with p and q
co-prime)
are chiral, as first noticed by Kneser. They are obtained from $S^3$ by
identifying points under the action of the cyclic group $C_p$
given by [12--13]:
\[
Z^1 \rightarrow \exp \left({{2 \pi i } \over p} \right) Z^1
\]
\[
Z^ 2 \rightarrow \exp \left({{2 \pi qi}  \over p} \right)Z^2
\]
Because the action of the cyclic group commutes with the
Hopf fibration the time-orientable Lorentz
metric described above descends to all of the lens spaces.

The topological classification of the lens spaces
depends on the bi-linear map:
\[
\lambda : H_1({\cal M}; {\Bbb Z}) \times  H_1({\cal M}; {\Bbb Z}) \rightarrow
{\Bbb Q}/{\Bbb Z}
\]
called the linking form defined on the first homology group
$H_1({\cal M}; {\Bbb Z}) \equiv {\Bbb Z} /p {\Bbb Z}$.
The linking form $\lambda $ changes sign under reversal of orientation.

Now the integral curve of the timelike vector field ${\bf V}$ gives
a generator $\gamma$
of $H_1({\cal M},{\Bbb Z})$ with linking invariant:
\[
\lambda ( \gamma, \gamma) = {q \over p}.
\]

The remaining elements $\alpha$ of  $H_1({\cal M},{\Bbb Z})$
are of the form $\alpha = x \gamma$, $x= 0,1,\dots p-1$. The bilinearity
of $\lambda(,)$
implies that
\[
\lambda (\alpha, \alpha) = x^2 { q \over p}.
\]
To exhibit a chiral lens space it suffices to find a pair of co-prime
natural numbers $(p,q)$ such that for no  $x=0,1,\dots p-1$ is it true that:
\[
x^2 { q \over p } = - { q \over p} \hskip 3cm ({\bmod} ~p).
\]
Thus $L_{3,1}$ is an example of a chiral three-dimensional spacetime.
On the other hand, the action of the cyclic group $C_p$
may or may not  commute with the reflection $\Gamma$.
If it does, then the action of $\Gamma$ will descend to the quotient
and then we can still reverse time.

We remark here that, consistent with our
general idea, the partition function  $Z({\cal M}_3)$
 for Witten's topological field theory is invariant under
all diffeomorphisms whether or not they are in the identity component
${\rm Diff} _0({\cal M})$, and obeys:
\[
Z({\overline {\cal M}}_3) = {\overline {Z({\cal M}_3)}}
\]
where $\overline {\cal M}$ is the same manifold as $\cal M$
but with the opposite orientation.
Thus for reversible manifolds it is real, and conversely if it is complex,
then the manifold must be chiral.

 Turning to four-dimensional manifolds: a standard example
of an irreversible four-manifold is
${\Bbb C}{\Bbb P}^2$. Notationally one distinguishes between
${\Bbb C}{\Bbb P}^2$ and ${\overline {{\Bbb C}{\Bbb P}}^2}$. The Euler
characters $\chi$
are the same, but the Hirzebruch signatures $\tau = b^+_2- b^-_2$
are opposite in sign:

\[
\tau ({\Bbb C}{\Bbb P}^2) = 1= - \tau ({\overline {\Bbb C}{\Bbb P}^2}).
\]
Quite generally, a four-manifold with non-vanishing
Hirzebruch signature cannot admit an orientation-reversing diffeomorphism.
Now  consider, for example  the connected sum of $K3$ with $12$ copies of
$S^1 \times S^3$. This has vanishing Euler characteristic and signature $16$.
It therefore
admits no total orientation-reversing diffeomorphism but the Lorentz
structure  $g^L$ is homotopic to the time-reversed Lorentz structure.
We do not know, however,  whether there exist diffeomorphisms (connected to the
identity or not)
which will produce this time-reversal.\\

\section {Generalized Taub-NUT Spacetimes}

The four-dimensional Taub-NUT solution
of Einstein's vacuum equations has provided many examples of the
possible exotic behaviour of Lorentzian metrics.
In this section we provide a family
of higher-dimensional examples, based on some work by Bais
and Batenberg [3] on the associated Riemannian metrics,
 which serve to illustrate
our general results.

Suppose
$\{{\cal B}, g^{\cal B}, \omega ^{\cal B} \}$ is a $2p$-dimensional
 Einstein-K\"ahler manifold with
 K\"ahler form $\omega ^{\cal B}$ which obeys the
Dirac quantization condition, i.e., it represents an integral class
\[
\left[{ 1 \over { 2 \pi} }   \omega ^{\cal B} \right] \in H_2( {\cal B}; {\Bbb
Z})
\]

Then $\omega ^{\cal B} $ may be thought of as the curvature of an $S^1$
bundle over ${\cal B}$. Let
\[
e^0 = d t + A
\]
where $0 \le t <2 \pi$ be a coordinate on the $S^1$ fibre and $A$
the connection such that:
\[
dA= \omega ^{\cal B}
\]

Then the $(2p+2)$-dimensional time-orientable Lorentzian metric

\[
F^{-1} (r) dr^2  + ( r^2 + N^2) g^{\cal B} - 4N^2 F(r) e^0 \otimes e^0
\]
is Ricci flat, provided
\[
F(r) = {r \over { (r^2 +N^2)^p}} \int ^r (s^2+ N^2)^p {{ds} \over {s^2}}
\]
The function $F(r)$ contains two arbitrary constants, the generalized
\lq NUT \rq ~charge $N$ and an arbitary constant of integration.
If $p=1$ then $\{{\cal B}, g^{\cal B}, \omega ^{\cal B} \}$ is
${\Bbb C}{\Bbb P}^1 \equiv
S^2$, the $S^1$ bundle is $S^3$ and we recover the usual
Taub-NUT solution. Indeed, when $p = 1$ we have
\[
F(r) = {r \over { (r^2 +N^2)}}(r -{N^2 \over {r}} -2m)
\]
where m is a constant of integration.  One now recovers the
usual Taub-NUT metric [5] with ${\rm A} = {\cos\theta}{\rm d}\phi$,
${\rm t} = \psi$ and ${\rm r} = {\rm t}$.

For higher values of $p$ one finds that
\begin{eqnarray*}
F(r) &=& {1 \over {(r^2 + N^2)}}( {r^2p \over (2p -1)} + {p{N^2}{r^2p -2} \over
(2p -3)}
+ ... -N^2p - 2mr) \\
&=& {1 \over {(r^2 + N^2)}}({r^2 \over (2p -1)}P(r) - N^2p - 2mr)
\end{eqnarray*}
where $P(r)$ is a polynomial of degree $2(p -1)$ containing only even powers
of ${\rm r}$, all of whose coefficients are positive.  It follows that the
numerator of $F({\rm r})$ has just two real roots.

In these higher $p$ generalisations we
can choose $\{{\cal B}, g^{\cal B}, \omega ^{\cal B} \}$ to be
${\Bbb C}{\Bbb P}^p$ and then the $S^1$ bundle becomes $S^{2p+1}$ with
its standard Hopf fibration. In this case, the isometry group
of the spacetime is $U(p+1)$ which acts transitively on $S^{2p+1}$
and contains a $U(1)$ factor acting as time translations.
The group acting on the base $\cal B$ is $SU(2p)/{\Bbb Z}_{2p}$.

The resulting $(2 p +2)$-dimensional
spacetime may be thought of as a time-orientable
two-plane bundle over $\cal B$
carrying an $SO(2)$-invariant Lorentzian metric on the fibres with  local
coordinates $t,r$. Its structure is independent of the particular
metric on the base $\cal B$.  Because the numerator of $F({\rm r})$
has only two real roots the structure is qualitatively the same
as that of the usual four-dimensional Taub-NUT case.  In particular, the
Penrose diagram is the same as that shown on page 177 of [5].  Note that
if the constant of integration $m$ is chosen to vanish, the metric has
an additional discrete isometry ${\rm r} \rightarrow {\rm -r}$,
interchanging different asymptotic regions.

{}From the point of view of this paper, we are interested in whether one
can find a diffeomorphism which reverses the time-orientation.
If we consider the case when ${\cal B} \equiv {\Bbb C}{\Bbb P}^p$,
and we confine ourselves to diffeomorphisms keeping the coordinate
$r$ fixed, then we are in the same position as above in our discussion
of the odd-dimensional spheres $S^{2p+1}$. Thus if $p$ is odd,
we can and if $p$ is even, we cannot reverse the sense of time
by means of a diffeomorphism in the identity component
${\rm Diff} _0( {\cal M})$.  Presumably this means that there
is no invariant significance in the sign of the NUT charge
$N$ if $p$ is odd but there may be if $p$ is even.
{\it Presumably therefore if $p$ is odd, as it is in the usual
four-dimensional case, then a Taub-NUT solution
should be considered as its own anti-particle}. \\

\section {Time-Reversal for Dynamical Systems}

The topological ideas about
time reversal discussed in this lecture may be applied
in a different but related  context.
Suppose we have a finite-dimensional
autonomous dynamical system with a compact phase space. That is, we have a
symplectic manifold $\{ {\cal M}_{2r}, \omega \}$
with symplectic form $\omega$ and Hamiltonian vector field
\[
{\bf H} = \omega ^{-1} d H .
\]
Now time reversal is an {\it anti-symplectic involution}
\[
f: {\cal M }_{2r}  \rightarrow {\cal M}_{2r}; \> f^2 = {\rm id}
\]
such that
\[
f^* \omega = - \omega.
\]
and
\[
f^* H =H
\]
and therefore
\[
f_* {\bf H} = - {\bf H}
\]
The standard (non-compact) example is of course ${\Bbb R}^{2r} \equiv T^* (
 {\Bbb R} ^r ) $ for which $f: ({\bf q}, {\bf p}) \rightarrow ({\bf q}, -{\bf
p} ) $.
To get a compact example,
one may replace ${\Bbb R}^r$ by any configuration space
manifold $\cal Q$. If $\cal Q$ were compact and we had a Hamiltonian
action of some symmetry group $G$, we might pass to the symplectic
quotient which might be compact.

Since the $r$-th power:
\[
\omega \wedge \dots \wedge \omega
\]
defines a volume form on ${\cal M}_{2r}$,
time reversal is orientation-reversing if $r$ is odd and orientation-
preserving if $r$ is even.

Thus if $r$ is odd, it cannot live in the identity
component ${\rm Diff}_0({\cal M}_{2r})$. Therefore if $r$ is odd
and ${\cal M}_{2r}$ is irrevesible, then no such $f$ can exist.
{\it Thus if such an $\{ {\cal M}_{2r}, \omega \}$ exists, it would
mean that {\it no} dynamical sytem on this
phase space, whatever its Hamiltonian function, could be invariant under time
reversal!}\\

\section {Some References}

\noindent [1] P.C.W. Davies, {\it Nat. Phys. Sci.} {\bf 240}, 3--5 (1972)\\

\noindent [2] T. Gold, {\it Amer. Journ. Phys.}, {\bf 30}, 403 (1962)\\

\noindent [3] S. Bais and P. Batenberg, {\it Nucl. Phys.} {\bf B253}, 162
(1985)\\

\noindent [4] M.C. Albrow, {\it Nat. Phys. Sci.} {\bf 241}, 56--57 (1973)\\

\noindent [5] S.W. Hawking and G.F.R. Ellis, {\it The large scale structure
of spacetime}, CUP Cambridge (1973)\\

\noindent [6] A. Chamblin, {\it Comm. Math. Phys.} {\bf 164}, No. 1, pgs.
65--87 (1994).\\

\noindent [7] B.S. Kay, {\it Rev. Math. Phys., Special Issue}, 167--195
(1992)\\

\noindent [8] K.K. Lee, {\it Gen. Rel. Grav.} {\bf 4}, 421--433 (1973)\\

\noindent [9] K.K. Lee, {\it Gen. Rel. Grav.} {\bf 5}, 239--242 (1973)\\

\noindent [10] K.T. Chen, {\it Proc. Amer. Math. Soc.} {\bf 26}, 196 (1970)\\

\noindent [11] P.A. Smith, {\it Ann. Math.} {\bf 37}, 526 (1936)\\

\noindent [12] J. Hempel, {\it Ann. Math. Studies} {\bf 86}, Princeton Univ.
Press,
Princeton (1976)\\

\noindent [13] H. Seifert and W. Threlfall, {\it A Textbook of Topology},
Academic Press (1980)\\

\noindent [14] E. Thomas, {\it Bull. Amer. Math. Soc.} {\bf 75}, 643--668
(1969)\\

\noindent [15] M.F. Atiyah, {\it Arbeitsgemeinschaft f{\"u}r Forschung des
Landes
Nordrhein-Westfalen}, Heft 200\\

\noindent [16] G.W. Gibbons and H.J. Pohle, {\it Nucl. Phys. B} {\bf 410},
117--142
(1993)\\

\noindent [17] G.W. Gibbons, {\it Int. Journ. of Mod. Physics D} {\bf 3}, No. 1
(1994)\\

\noindent [18] A.D. Sakharov, {\it Sov. Phys. JETP} {\bf 60}, 214 (1984)\\

\noindent [19] A. Chamblin and G.W. Gibbons, {\it Class. Quant. Grav.} {\bf
12}, No.
9, 2243 (1995)\\

\noindent [20] J. Friedman, {\it Class. Quant. Grav.} {\bf 12}, No. 9 2231
(1995)\\

\noindent [21] J. Friedman and A. Higuchi, gr-qc \# 9505035\\

\noindent [22] M. Gell-Mann and J. Hartle, {\it Complexity, Entropy, and the
Physics of Information, SFI Studies in the Science of Complexity, Vol. VIII},
pages 425-458.  Addison-Wesley, Reading (1990) \\

\noindent [23] A. Vilenkin, {\it Phys. Rev. D} {\bf 27}, 2848 (1983)\\

\noindent [24] J. Milnor and D. Husemoller, {\it Symmetric Bilinear Forms},
Springer-Verlag (1973)\\

\noindent [25] G.W. Gibbons, {\it Nucl. Phys. B} {\bf 291}, 497 (1986)\\

\noindent [26] W. Massey, {\it Algebraic Topology, An Introduction},
Harcourt, Brace and World (1967)\\

\noindent [27] E. Witten, {\it Nucl. Phys. B} {\bf 195}, 481 (1982)\\

\noindent [28] F. Hirzebruch and H. Hopf, {\it Math. Ann.} {\bf 136}, 156--172
(1958)\\

\end{document}